\documentstyle[12pt]{article}
\begin{document}
\begin{titlepage}
\hoffset = -1truecm
\voffset = -2truecm
\title{\bf
U--duality and Simplectic Formulation of Dilaton--Axion Gravity
 }
\author{
{\bf
D.V. Gal'tsov}
\thanks{
Permanent Address:
Department of Theoretical Physics,  Moscow State University, Moscow 119899,
Russia, e--mail:
galtsov@grg.phys.msu.su}
\\
\normalsize International Centre for Theoretical Physics, Trieste 34100, {\bf
Italy}\\
  \and
{\bf  O.V.  Kechkin}\\
\normalsize  Department of Theoretical Physics,  Moscow State University,\\
\normalsize Moscow 119899, {\bf Russia}
 }
\normalsize \date{30th June 1995}
\maketitle
\begin{abstract}
We study a bosonic four--dimensional effective action
corresponding to the heterotic string compactified on a 6--torus
(dilaton--axion gravity with one vector field) on a curved space--time
manifold possessing a time--like Killing vector field.
Previously an existence of the $SO(2,3)\sim Sp(4, R)$ global
symmetry ($U$--duality)  as well as the symmetric space property
of the corresponding $\sigma$--model have been established following
Neugebauer and Kramer approach.
Here we present an explicit form of the $Sp(4, R)$ generators in terms
of coset
variables and construct a representation of the coset in terms of
the physical target space
coordinates.  Complex symmetric $2\times 2$ matrix  $Z$
(``matrix  dilaton --axion'')  is then introduced for which $U$--duality
takes the matrix valued $SL(2, R)$ form. In terms of this matrix the theory
is further presented as a K\"ahler $\sigma$--model. This leads to a more
concise $2\times 2$ formulation
which opens  new ways to construct exact classical solutions. New solution
(corresponding to constant ${\rm Im} Z$ ) is obtained which
describes the  system of point massless magnetic monopoles
endowed with axion charges equal to minus monopole charges.
In such a system mutual magnetic repulsion
is exactly balanced by axion  attraction so that the resulting
space--time is locally flat but possesses multiple Taub--NUT singularities.
\vskip5mm
\noindent
PASC number(s): 97.60.Lf, 04.60.+n, 11.17.+y
\end{abstract}
\end{titlepage}
\newpage
\section{Introduction}
\renewcommand{\theequation}{1.\arabic{equation}}
Duality symmetries relating different backgrounds which define
essentially the same quantum conformal field theories play an important role
in the theory of superstrings (for a review see e.g. \cite{gpr},
\cite{al}).
Recently a considerable progress has been achieved in understanding dualities
of toroidally compactified heterotic string \cite{nr}.
One such symmetry is T--duality  relating different backgrounds
whoose geometries admit an Abelian isometry group \cite{odp}.
In the zero--slope limit the relevant effective action in four dimensions
contains a space--time metric,
a Kalb--Ramond antisymmetric tensor field,
a dilaton and moduli fields forming some coset $\sigma$--model,
as well as a number of vector fields. At a generic point of the moduli
space vector fields should form an Abelian multiplet of dimension $p+d$,
where $p$ is a nubmer of initial vector fields in the $D$--dimensional theory
and $d$ is a number of compactified dimensions, or, in other terms, a rank
of an Abelian space--time isometry group
(for $D=10$ heterotic string $d=6,\; p=16$).
Classical T--duality of the toroidally compactified heterotic string
is then the global non--compact group
$SO(d, d+p)$ \cite{odp}, and it is conjectured that its discrete subgroup
$SO(d, d+p, Z)$ is an exact symmetry for the heterotic string.

In four (and lower) dimensions there is another notable symmetry, S--duality
\cite{sd}, which emerges essentially due to the fact that in four
dimensions an antisymmetric three--index field is dual to a (pseudo--)
scalar. Together with a dilaton (or an effective dilaton) these two scalars
form a coset $SL(2, R)/SO(2)$. The $SL(2, R)$--symmetry also persists
when the system is coupled to $U(1)$ vector fields. It includes
an interchange of electric and magnetic components of vector fields as
well as an inversion operation relating strong and weak coupling regimes.
It is beleived that the discrete subgroup $SL(2, Z)$ is a symmetry
of the full quantum heterotic string theory.
Together these dualities in 4 dimensions form a symmetry group
$T\times S = SO(d, d+p)\times SL(2, R)$
classically ($SO(d, d+p, Z)\times SL(2, Z)$
at the quantum level) \cite{s4}.

 It can be noted that the number of initial dimensions
$D$ does not enter as a parameter identifying these symmetry groups.
Although, of course,
$D$ should be equal to ten for an overall consistency of the quantum
heterotic string theory, and a number of primary vector fields
should be $p=16$ implying that $d=6,\; p+d=22$, the corresponding classical
field models exibiting $T$ and $S$ dualities may be considered
for different $D,\; d,$ and $p$. The simplest model of this kind,
which is often called dilaton--axion gravity, is formulated initially in
 $D=4$ where it
contains one vector field ($p=1$), a dilaton and an axion. It can be regarded
as the bosonic part of the toroidally compactified heterotic string effective
action with only one vector field and no moduli fields excited. Its symmetries
can easily be uplifted to the full ten--dimensional theory and they are
essential for an understanding of the space--time nature of string dualities.
At the same time, this theory may be considered as a generalization of the
Einstein--Maxwell theory, for which hidden symmetries arising from
dimesional reduction are well understood.

With this motivation we investigated four--dimensional dilaton--axion
gravity on the class of space--times admitting an existence
of a non--null Killing vector field, i.e. $D=4, d=p=1$\cite{gk}.
In this case T--duality is $SO(1,2)$ and one should
expect to have the full symmetry under a six--parametric group
$SO(1,2)\times SL(2,R)$. However, as it was shown in \cite{gk},
the full symmetry group in this case is {\em ten--parametric} and includes
both $T$ ans $S$ dualities as subgroups. This symmetry enhancement
was interpreted in \cite{gk} as manifestation of the unification of
$T$ and $S$ dualities in three dimensions.
 Remarkably, T--duality part of the full group turned out to correspond
to Ehlers--Harrison--type transformations known previously in the
Einstein--Maxwell theory \cite{eh}.
The group was then identified with the $SO(2,3)$ \cite{g} what meant
that we had an enhanced symmetry under a group
\begin{equation}
G_{d,p}=SO(d+1, d+p+1),
\end{equation}
for which a name of ``U--duality''was coined recently.

Soon later similar enhancement of duality symmetries in three dimensions
was reported by Sen \cite{s3} who investigated in detail the full
heterotic string case.
The resulting three--dimensional effective
theory (D=10, d=7, p=16) is supergravity with
eight local supersymmetries. Such theory was previously
studied by Marcus and Schwarz \cite{ms} who have found the
global symmetry under $SO(8, 24)$ which is just $G_{7, 16}$ in (1.1). Sen
presented an explicit dimensional reduction in terms of the heterotic string
variables including moduli fields and  argued that the discrete
subgroup of this group should be the  symmetry of the full quantum theory.
More general discussion of the unification of string dualities in dimensions
lower than four was given by Hull and Townsend \cite{ht} and recently by
Schwarz \cite{sh1} and Witten \cite{wi}.

Even more striking symmetry enhancement may arise when the theory is further
reduced to two dimensions. If in three dimensions the theory can be presented
as gravity coupled $\sigma$--model on a {\em symmetric} target space, its
two--dimensional reduction is known to be completely integrable \cite{maz},
what
entails infinite dimensional symmetries. The symmetric space property of the
three--dimensional $\sigma$--model target space seems to
be essential for further two--dimensional integrability of the system.
A notable counterexample is provided by the purely dilatonic theory
(with a vector field), which
possesses the symmetric space and two--dimensional integrability properties
only for exceptional values of the dilaton coupling constant \cite{gg}.
For pure dilaton--axion  gravity coupled $D=4$ system
(without vector fields) the symmetric space property
in three--dimensions follows directly
>from the coset nature of the dilaton--axion pair and similar property of the
vacuum Einstein gravity. Two--dimensional reduction of this theory
 was discussed by Bakas \cite{bk}. He showed
that, similarly to the vacuum Einstein equations, an infinite dimensional
symmetry (Geroch group) then comes into play. Geroch group is generated by the
$SO(2,2)$ current algebra and is essentially the loop extension
$\widehat{SO(2,2)}$. The same theory was also discussed
by Maharana \cite{mah}.

Dilaton--axion gravity including a vector field (D=4, p=1) was studied
by one of the authors \cite{g} . It was shown that the three--dimensional
$\sigma$--model target space is a symmetric riemannian space indeed which can
be
identified with the coset $SO(2,3)/(SO(3)\times SO(2))$. In two
space--time dimensions this system was shown to admit a representation as a
modified chiral matrix model of the Belinskii--Zakharov type \cite{bz}.
This entails an infinite symmetry under the loop group
$\widehat{SO(2,3)}$ generated by the corresponding matrix current algebra
which is similar to the Kinnersley--Chitre group for the Einstein--Maxwell
equations \cite{ki}.

The full heterotic string two--dimensional case was studied recently
by Sen \cite{s2},
more general discussion was given by Schwarz \cite{sh2} (see also  recent
papers \cite{alk}).
The corresponding infinite--dimensional loop group is then
$\widehat{SO(8, 24)}$ as could be expected.

Here we continue an investigation of the $D=4, d=p=1$ theory
(dilaton--axion gravity with one vector field reduced to three dimensions).
This theory possessing a U--duality group $SO(2,3)$
has another interesting feature which has been
noted in \cite{g}: it admits an alternative symplectic formulation
due to isomorphism
\[
SO(2,3)\sim Sp(4,R).
\]
This opens a way to construct a more concise representation of the system in
terms
of $4\times 4$ matrices instead of $5\times 5$ ones. The symplectic group has
also
an advantage to be directly related to the charge
quantization \cite{gz}. The purpose of the present paper is to develop the
simplectic formalism in detail. We find an elegant description
of the system in terms of $2\times 2$ complex
symmetric matrix (``matrix dilaton--axion'') which under action of the
$Sp(4, R)$ undergoes a matrix--valued $SL(2, R)$ transformation.
It will be shown that in terms of components of this complex matrix we deal
with the K\"ahler $\sigma$--model.

There is yet another motivation for studying four--dimensional dilaton--axion
gravity  related to the investigation of non--flat
four--dimensional string backgrounds (see e.g. \cite {hts}) and references
therein). Special interest to such backgrounds is related to the
quantum black hole problem. Some particular four--dimensional black
hole solutions to the dilaton--axion gravity were successfully interpreted
in terms of gauged WZW models \cite{wzw}. One of the main problems
of this approach is to find a way to construct more general exact string
backgrounds. Study of classical solutions at present seems to be
a necessary step to this goal. The most general seven--parametric
family of classical four--dimensional black hole solutions (generalizing
Gibbons--Garfinkle--Strominger solution \cite{gi})
to the heterotic string theory which included both electric and
magnetic charges, rotation and NUT parameters was found in
\cite{gk} using $d=p=1\;U$--duality transformations.
Extremal members of this family received
recently a WZW interpretation in the work of Johnson and Myers
\cite{jm}. (Note also a ten--dimensional classical embedding of the above
solutions \cite{sbh}). Our present formulation
opens a way to apply more powerful techniques in order to construct exact
classical solutions to dilaton--axion gravity
such as an inverse scattering transform method and Backlund transformations.
Although we do not consider their direct application here, we derive
some very simple (but apparently previously unknown) solution which is directly
suggested by the new formalism.

Our treatment essentially follows Neugebauer and Kramer approach
developed long ago for the Einstein--Maxwell theory \cite{nk}. Contrary
to the standard dimensional reduction in the matrix form,
this is a local technique based on the existence of a
$\sigma$--model representation of three--dimensional gravity. Isometries
of the target space are first derived in infinitesimal
form by solving the corresponding Killing equations. This was done
earlier in \cite{nk} for the Einstein--Maxwell equations.

The plan of the paper is as follows. We start (Sec. 2)
with the brief outline of
the derivation  of three--dimensional $\sigma$--model (more detailes can
be found in \cite{ggk}). In Sec. 3 Killing vectors exibiting symmetries
of the target space are presented and a matrix representation is obtained
for their algebra. Sec. 4 is devoted to the construction of the coset
$Sp(4,R)/U(2)$ in terms of the target space variables. Several alternative
formulations of the theory in terms of $2\times 2$ matrices are also given,
in particular, a K\"ahler representation. In Sec. 5 we
construct new solution describing a system of magnetic charges
endowed with axion charges in static equilibrium. The corresponding
geometry is the massless multi--Taub--NUT space--time. In the Appendix some
detailes of the derivation of the coset in terms of three complex
variables are given.
\section{ Three--dimensional $\sigma$--model}
\renewcommand{\theequation}{2.\arabic{equation}}
\setcounter{equation}{0}
We start with the $D=4,\; p=1$ bosonic sector of the heterotic string
effective action on  curved space--time manifold
\begin{equation}
S=\frac{1}{16\pi}\int \left\{-R+2\partial_\mu\phi\partial^\mu\phi +
\frac{1}{2} e^{4\phi}
{\partial_\mu}\kappa\partial^\mu\kappa
-e^{-2\phi}F_{\mu\nu}F^{\mu\nu}-\kappa F_{\mu\nu}{\tilde F}^{\mu\nu}\right\}
\sqrt{-g}d^4x,
\end{equation}
where ${\tilde F}^{\mu\nu}=\frac{1}{2}E^{\mu\nu\lambda\tau}F_{\lambda\tau}$,
and $\phi,\; \kappa$
are dilaton and axion fields . To reduce the system to three dimensions
we assume that space--time admits a non--null
Killing vector (here choosen to be timelike).
Then 4--dimensional metric can be presented
in terms of a three--metric $h_{ij}$, a rotation one--form
$\omega_i,\; (i, j=1, 2, 3)$
and a three--dimensional conformal factor $f$ depending only
on the 3--space coordinates $x^i$:
\begin{equation}
ds^2=g_{\mu\nu}dx^\mu dx^\nu=f(dt-\omega_idx^i)^2-\frac{1}{f}h_{ij}dx^idx^j.
\end{equation}

The subsequent derivation of a three--dimensional $\sigma$--model
representation is rather standard (for more details see \cite{ggk})
and essentially follows Israel and Wilson \cite{iw}.
Four--dimensional vector field in three dimensions is represented by two
scalars. In our case of timelike Killing symmetry they have a meaning of
electric $v$
and magnetic $u$ potentials which can be introduced via relations
\begin{equation}
F_{i0}=\frac{1}{\sqrt{2}}\partial_iv,
\end{equation}
\begin{equation}
e^{-2\phi}F^{ij}+\kappa {\tilde F}^{ij}=\frac{f}{\sqrt{2h}}\epsilon^{ijk}
\partial_ku
\end{equation}
solving a spatial part of the  modified Maxwell equations and
Bianchi identity. A rotation one--form $\omega_i$ in (2.2)
also reduces to the scalar --- a twist potential
 $\chi$ :
\begin{equation}
\tau^i=-f^2\frac{\epsilon^{ijk}}{\sqrt{h}}\partial_j\omega_k,\;\;
\tau_i=\partial_i\chi +v\partial_iu-u\partial_iv.
\end{equation}

Therefore in three dimensions we have three pairs of scalar variables:
the first $(f,\;\chi)$ is inherited from the four--dimensional metric,
the second $(v,\;u)$ --- from vector field, the last is the
dilaton--axion pair $(\phi, \kappa)$.
A remarkable feature of the four dimensional gravitationally coupled
system of $U(1)$ vector fields interacting with scalars representing
some coset is that in three
dimensions they form a gravity coupled sigma--model \cite{bgm}. Most notable
examples are provided by the Einstein--Maxwell system \cite{nk}, and
multidimensional
vacuum Einstein equations compactified to four dimensions \cite{ma}. In our
case it can be  checked directly that the equations for
$f, \chi, v, u, \phi, \kappa$ may be
obtained by variation of the following three--dimensional action
\[
S=\int\Bigl\{{\cal R}-\frac{1}{2f^2}[(\nabla f)^2+(\nabla\chi +
v\nabla u-u\nabla v)^2]-2(\nabla\Phi )^2-
\]
\begin{equation}
-\frac{1}{2}e^{4\phi}(\nabla\kappa )^2+\frac{1}{f}\left[e^{2\phi}
(\nabla u-\kappa\nabla v)^2+
e^{-2\phi}(\nabla v)^2\right]\Bigr\}\sqrt{h}d^3x,
\end{equation}
where ${\cal R}\equiv {\cal R}_i^i$ and $\bf\nabla$ stands for 3--dimensional
covariant derivative.

The remaining (spatial) Einstein equations may be  regarded as
3--dimensional Einstein equations
for the metric $h_{ij}$
 \[
{\cal R}_{ij}=\frac{1}{2f^2}(f_{,i}f_{,j}+\tau_i\tau_j)+2\phi_{,i}
\phi_{j}+\frac{1}{2}e^{4\phi}\kappa_{,i}\kappa_{,j}-
\]
\begin{equation}
-\frac{1}{f}\left[e^{-2\phi}v_{,i}v_{,j}+e^{2\phi}(u_{,i}-\kappa v_{,i})
(u_{,j}-\kappa v_{,j})\right],
\end{equation}
where the source term is derivable from the same action as the 3--dimensional
energy momentum tensor
(here ${\cal R}_{ij}$ is  three--dimensional Ricci tensor).

 This action  can be rewritten as the gravity coupled three--dimensional
$\sigma$--model
\begin{equation}
S=\int \left({\cal R}-{\cal G}_{AB}\partial_i\varphi^A\partial_j\varphi^B
h^{ij}\right)\sqrt{h}d^3x,
\end{equation}
where $\varphi^A=(f,\; \chi,\; v,\;u,\;\kappa,\; \phi,)\; A=1,..., 6$.
The corresponding target space metric reads
\begin{eqnarray}
{dl}^2 & = & \frac{1}{2}f^{-2}[{df}^2+
(d\chi +vdu-udv)^2]-f^{-1}[e^{2\phi }{(du-\kappa dv)}^2
+e^{-2\phi }{dv}^2] \nonumber \\
& + & 2{d\phi }^2+
\frac{1}{2}e^{4\phi }{d\kappa }^2.
\end{eqnarray}
This expression generalizes Neugebauer and Kramer potential space metric
found for the stationary Einstein--Maxwell system in 1969 \cite{nk}.
It is worth noting that the present $\sigma$--model does not reduce to the
Einstein--Maxwell one if $\kappa=\phi=0$, because the equations for
a dilaton  and an axion  generate constraints $F^2=F\tilde F =0$. Hence,
generally
the solutions
of the Einstein--Maxwell theory with one Killing symmetry are not related
to solutions of the present theory by target space isometries (except for
the case $F^2=F\tilde F =0$). From the other hand, one can consistently set
in the present $\sigma$--model $f=\chi=\phi=\kappa=0$ reducing it to the
Einstein
vacuum sigma--model. Therefore all solutions to the vacuum Einstein
equations with one non--null Killing symmetry
are related to some solutions of the
system in question by target space isometries. This fact was used in \cite{gk}
to generate the most general black hole solutions in dilaton--axion gravity.
These isometries form  $D=4,\;d=p=1\;U$--duality group.

\section{$Sp(4, R)$ symmetry}
\renewcommand{\theequation}{3.\arabic{equation}}
\setcounter{equation}{0}
In the previous work \cite{gk} it was found that the target space (2.9)
possesses a 10--parametric isometry
group containing $T$--duality (Ehlers--Harrison transformations),
together with the $SL(2,R)$
electric--magnetic duality, gauge a scale transformations. This isometry group
was
subsequently identified with $SO(2,3)$ isomorphic to $Sp(4, R)$ \cite{g}.
Here we intend to give more detailed description of this symmetry.
Standard procedure to reveal hidden symmetries in supergravity and string
theory consists in direct dimensional reduction in the matrix form.
In \cite{gk} we followed
another approach \cite{nk} starting from the infinitesimal isometries of
the target space
and then deriving finite group transformations by means of integration.

 Killing vectors, satisfying the equations
\begin{equation}
K_{A;B}+K_{B;A}=0,
\end{equation}
where covariant derivatives refer to the target space metric (2.9) read as
follows. Graviational gauge transformation $\chi\rightarrow \chi+const$ and
electromagnetic gauge transformations are generated by Killing vectors
\begin{equation}
 {K}_g=\partial_{\chi },
\end{equation}
\begin{equation}
 {K}_m=\partial_u+v\partial_{\chi },
\end{equation}
\begin{equation}
 {K}_e=\partial_v-u\partial_{\chi }.
\end{equation}
Electric--magnetic duality subalgebra $sl(2, R)$ consists of two
rotations
\begin{equation}
 {K}_{d_1}=\partial_{\kappa }+v\partial_u,
\end{equation}
\begin{equation}
 {K}_{d_2}=(e^{-4\phi }-{\kappa }^2)\partial_{\kappa }+
\kappa \partial_ \phi +u\partial_v,
\end{equation}
and a dilaton shift accompanied by suitable transformations of electromagnetic
potentials
and of an axion
\begin{equation}
 {K}_{d_3}=\partial_\phi -2\kappa \partial_{\kappa }+v\partial_v-u\partial_u.
\end{equation}
They satisfy the following commutation relations
\begin{equation}
\left[K_{d_3},K_{d_1}\right]=2K_{d_3},\quad
\left[K_{d_3},K_{d_2}\right]=-2K_{d_2},\quad
\left[K_{d_1},K_{d_2}\right]=K_{d_3}.
\end{equation}
A scale transformation is generated by
\begin{equation}
 {K}_s=2f\partial_f+2\chi \partial_{\chi }+v\partial_v+u\partial_u .
\end{equation}
Form these seven Killing vectors six are rather obvious, while (3.6)
is easier to find firstly in the finite form (see \cite{gk}).

Essentially non--trivial part of the whole isometry group consists of
a conjugate pair of Harrison--type transformations which have been found by
solving the Killing equations (3.1) explicitly:
\begin{eqnarray}
 {K}_{H_1} &=& 2vf\partial_f+v\partial_\phi+2w\partial_{\kappa }+
(v^2+fe^{2\phi })\partial_v \nonumber \\
& + & (\chi+uv+\kappa fe^{2\phi })\partial_u+
(v\chi +wfe^{2\phi })\partial_{\chi },
\end{eqnarray}
\begin{eqnarray}
 {K}_{H_2} &=& 2uf\partial_f+(\kappa v-w)\partial_\phi+2(\kappa w +
 ve^{-4\phi })\partial_{\kappa }+
(uv-\chi +\kappa fe^{2\phi })\partial_v \nonumber \\
& + & (u^2+fe^{-2\phi }+{\kappa }^2fe^{2\phi })\partial_u+
(u\chi -vfe^{-2\phi }+\kappa wfe^{2\phi })\partial_{\chi },
\end{eqnarray}
where $w=u-\kappa v$. Their commutator corresponds to the Ehlers--type
transformation
\begin{equation}
\left[K_{H_1},K_{H_2}\right]=2K_{E},
\end{equation}
\begin{eqnarray}
{K}_{E} &=& 2f\chi \partial_f+wv\partial_\phi+(w^2-v^2e^{-4\phi
})\partial_{\kappa }+
(v\chi +wfe^{2\phi })\partial_v \nonumber \\
& + & (u\chi -vfe^{-2\phi }+\kappa wfe^{2\phi })\partial_u+
({\chi }^2-f^2+fv^2e^{-2\phi } +fw^2e^{2\phi })\partial_{\chi }.
\end{eqnarray}

Altogether these Killing vectors form a ten--dimensional algebra
$so(2,3)\sim sp(4, R)$  \cite{g}.
To prove this we construct below an explicit matrix representation of the
Killing algebra.
have to relate 10 above generators to $sp(4, R)$ algebra.
First we recall some properties of real symplectic groups.
An element of $Sp(2n, R)$ is a real
$2n\times 2n$ matrix $G$ preserving a skew ``metric'' $J$,
\begin{equation}
G^TJG=J,\quad
J=\left(\begin{array}{crc}
O&I\\
-I&O\\
\end{array}\right) .
\end{equation}
In view of the relation
\begin{equation}
J^2=-I,
\end{equation}
an inverse matrix is given by
\begin{equation}
G^{-1}=-JG^TJ,
\end{equation}
For infinitesimal transformations
\begin{equation}
G=I+A
\end{equation}
it follows from (3.14)
\begin{equation}
A^TJ+JA=0,
\end{equation}
and consequently one has the following representation of $sp(2n, R)$
algebra in terms of $n\times n$ matrices $B, C, D$:
\begin{equation}
A=\left(\begin{array}{crc}
B&C\\
D&-B^T\\
\end{array}\right),\quad C^T=C,\quad D^T=D.
\end{equation}

 Now consider the case $n=2$. Let $\sigma_\mu$
denote a set of four real $2\times 2$ matrices:
\begin{equation}
\sigma_0=I,\quad \sigma_1=\sigma_x,\quad \sigma_2=i\sigma_y,\quad
\sigma_3=\sigma_z,
\end{equation}
where $I$ is a $2\times 2$ unit and $\sigma_{x, y, z}$ are usual Pauli
matrices with $\sigma_z$ diagonal. It is convenient to arrange them according
the symmetry under a transposition:
\begin{equation}
\sigma_\mu=\left(\sigma_2, \sigma_a\right),\;\; a=0, 1, 3,\;\;
\sigma_a^T=\sigma_a, \;\;\sigma_2^T=-\sigma_2.
\end{equation}
 The full set of  ten $sp(4, R)$ generators satisfying (3.18) consists
of 6 symmetric
\begin{equation}
V_a=\frac{1}{2}\left(\begin{array}{crc}
0&\sigma_a\\
\sigma_a&0\\
\end{array}\right),\quad
W_a=\frac{1}{2}\left(\begin{array}{crc}
\sigma_a&0\\
0&-\sigma_a\\
\end{array}\right),\quad
\end{equation}
and 4 antisymmetric matrices
\begin{equation}
U_a=\frac{1}{2}\left(\begin{array}{crc}
0&\sigma_a\\
-\sigma_a&0\\
\end{array}\right),\quad
U_2=\frac{1}{2}\left(\begin{array}{crc}
\sigma_2&0\\
0&\sigma_2\\
\end{array}\right),\quad
\end{equation}
the latter corresponding to the $u(2)$ subalgebra.

Computing commutators of Killing vectors (3.2--13)
and using multiplication rules for  the above matrix generators one can find
the
following correspondence
\begin{eqnarray}
2K_{g}&\leftrightarrow &-U_0+U_2-V_1+W_3,\nonumber \\
2K_{E}&\leftrightarrow &-U_0+U_2+V_1-W_3 ,\nonumber \\
2K_{d_1}&\leftrightarrow & -U_0-U_2+V_1+W_3,\nonumber \\
2K_{d_2}&\leftrightarrow & U_0+U_2+V_1+W_3,\nonumber \\
2K_{s}&\leftrightarrow &-(V_3+W_1),\nonumber \\
2K_{d_3}&\leftrightarrow & V_3-W_1, \nonumber\\
2K_{e}&\leftrightarrow & U_1+V_0,\nonumber \\
2K_{H_1}&\leftrightarrow & U_1-V_0,\nonumber \\
2K_{m}&\leftrightarrow & W_0-U_3,\nonumber \\
2K_{H_2}&\leftrightarrow & -(W_0+U_3).
\end{eqnarray}
Thus, infinitesimal isometries of the target space (2.9) form a real
symplectic algebra $sp(4,R)$. This algebra is isomorphic to $so(2,3)$ which
is the relevant $U$--duality algebra in our case $d=p=1$ (see (1.1)).
 The corresponding finite transformations can be found in the previous
paper \cite{gk}.
\section{Coset representation}
\renewcommand{\theequation}{4.\arabic{equation}}
\setcounter{equation}{0}
An important feature of the target space metric (2.9) is that all covariant
derivatives of the corresponding Riemann tensor vanish,
$\nabla_A {R^B}_{CDE}=0$. It means that this space is a symmetric Riemannian
space. In fact this property can be hinted already from dimensional
grounds: the target space (2.9) is six--dimensional, while its isometry
group is ten--dimensional; this is sufficient to ensure the above property.
  Maximal compact subgroup of $Sp(4, R)$ is $U(2)$ generated
by ${\cal H} =(U_a, U_2)$, which is four--dimensional. The remaining
generators
form the coset $Sp(4, R)/U(2)$. Denoting these  six generators  as
${\cal V}=(V_a, W_a)$, one finds:
\begin{equation}
[{\cal V},{\cal V}]\in{\cal H},\;\;[{\cal V},{\cal H}]\in{\cal V},
\end{equation}
what proves the symmetric space property of the coset.

Our goal here is to construct an explicit representation of this coset
in terms of physical target space coordinates. Within our approach based
initially on the infinitesimal formulation, this can be done as follows.

For any $G\in Sp(4, R)$ one can perform a Gauss decomposition
\begin{equation}
G=G_LG_SG_R,
\end{equation}
where
\begin{equation}
G_R=\left(\begin{array}{crc}
I&R\\
O&I\\
\end{array}\right),\quad
G_S=
\left(\begin{array}{crc}
{S^T}^{-1}&O\\
O&S\\
\end{array}\right),\quad
G_L=
\left(\begin{array}{crc}
I&O\\
L&I\\
\end{array}\right),
\end{equation}
and $S, R, L$ are real  $ n\times n $ matrices, $R, L$ being symmetric,
$R^T=R,\;L^T=L$.
Useful relations are
\begin{equation}
G_{R_1}G_{R_2}=G_{R_1+R_2},\quad G_{S_1}G_{S_2}=G_{S_1S_2},\quad
G_{L_1}G_{L_2}=G_{L_1+L_2} ,
\end{equation}
so that for inverse matrices one has
$G_R^{-1}=G_{-R},\;\;G_S^{-1}=S_{-R},\;\;G_L^{-1}=G_{-L}$.
With this parametrization
\begin{equation}
G=\left(\begin{array}{crc}
{S^T}^{-1}&{S^T}^{-1}R\\
L{S^T}^{-1}&S+L{S^T}^{-1}R\\
\end{array}\right).
\end{equation}

A subset of symmetric symplectic matrices $M\in Sp(4,R),\;M^T=M$ constitutes a
coset
$Sp(4, R)/U(2)$. The corresponding Gauss decomposition reads
\begin{equation}
M={\cal Q}^T{\cal P}{\cal Q},
\end{equation}
 \begin{equation}
{\cal Q}=
\left(\begin{array}{crc}
I&Q\\
O&I\\
\end{array}\right),\quad
{\cal P}=
\left(\begin{array}{crc}
P^{-1}&O\\
O&P\\
\end{array}\right),
\end{equation}
where two real symmetric $2\times 2$ matrices are introduced
$ Q^T=Q,\; P^T=P$. Multiplying  matrices we obtain
\begin{equation}
M=\left(\begin{array}{crc}
P^{-1}&P^{-1}Q\\
QP^{-1}&P+QP^{-1}Q\\
\end{array}\right).
\end{equation}

We find useful to combine $P$ and $Q$ into one complex symmetric matrix
 \begin{equation}
Z=Q+iP,
\end{equation}
which can be called ``matrix dilaton--axion''  for the following reason.
Consider a transformation of the coset under an action of  $G\in Sp(4, R)$ :
\begin{equation}
M\rightarrow G^TMG.
\end{equation}
Using the above formulas one can show that for $G=G_R$ :
\begin{equation}
Z \rightarrow Z+R,
\end{equation}
for $G=G_L$ :
\begin{equation}
Z^{-1} \rightarrow Z^{-1}-L,
\end{equation}
 while for $G=G_S$ :
\begin{equation}
Z \rightarrow S^TZS.
\end{equation}
These transformations are precisely the matrix analog of the
$SL(2, R)$ transformations which hold for the usual scalar
dilaton--axion field
\begin{equation}
z=\kappa +i e^{-2\phi}
\end{equation}
under S--duality generated by $K_{d_1},\;K_{d_2},\;K_{d_3}$ (see [7]).

Parametrizing $Z$ by three complex variables
\begin{equation}
Z=\left(\begin{array}{crc}
z_1&z_3\\
z_3&z_2\\
\end{array}\right),
\end{equation}
one can easily find infinitesimal generators of $Sp(4, R)$ in terms of $z_1,
z_2, z_3$.
Under infinitesimal
$R$--transformation, $R=I+\delta R$, where
\begin{equation}
\delta R=
\left(\begin{array}{crc}
\delta r_1&\delta r_3\\
\delta r_3&\delta r_2 \\
\end{array}\right),
\end{equation}
$Z$ transforms as
\begin{equation}
Z \rightarrow Z+\delta R.
\end{equation}
In the space of holomorphic functions of the complex variables
$z_i,\; i=1, 2, 3$ these transformations are generated by
\begin{equation}
R_i=\partial _{z_i}\equiv\partial _i .
\end{equation}

Similarly, for infinitesimal $S$--transformation, $S=I+\delta S$,
\begin{equation}
\delta S=
\left(\begin{array}{crc}
\delta s_1&\delta s_3\\
\delta s_4&\delta s_2 \\
\end{array}\right),
\end{equation}
one has
\begin{equation}
Z \rightarrow Z+{\delta S}^T Z +Z\delta S,
\end{equation}
and consequently
\[
S_1=2z_1\partial_1+z_3\partial_3,\quad
S_2=2z_2\partial_2+z_3\partial_3,
\]
\begin{equation}
S_3=2z_3\partial_2+z_1\partial_3,\quad
S_4=2z_3\partial_1+z_2\partial_3.
\end{equation}
Finally, for infinitesimal $L$--transformation, $L=I+\delta L$,
\begin{equation}
\delta L=
\left(\begin{array}{crc}
\delta l_1&\delta l_3\\
\delta l_3&\delta l_2 \\
\end{array}\right),
\end{equation}
one finds
\begin{equation}
Z^{-1}\rightarrow Z^{-1}-\delta L,
\end{equation}
so that the corresponding differential operators are
\[
L_1={z_1}^2\partial_1+{z_3}^2\partial_2+z_1z_3\partial_3,\quad
L_2={z_2}^2\partial_2+{z_3}^2\partial_1+z_2z_3\partial_3,
\]
\begin{equation}
L_3=2z_1z_3\partial_1+2z_2z_3\partial_2+(z_1z_2+{z_3}^2)\partial_3.
\end{equation}

Now, it is a simple matter to find the correspondence between the matrices
(3.22-23) and generators in terms of $z_i$ by constructing $4\times 4$ matrix
representation of (4.18), (4.21)  and (4.24). The result is
 \begin{eqnarray}
&&2U_0=R_1+R_2+L_1+L_2 ,\nonumber \\
&&2V_0=R_1+R_2-L_1-L_2 ,\nonumber \\
&&2U_3=R_1-R_2+L_1-L_2 ,\nonumber \\
&&2V_3=R_1-R_2-L_1+L_2 ,\nonumber \\
&&2U_1=R_1+L_3 ,\nonumber \\
&&2V_1=R_1-L_3 ,\nonumber \\
&&2W_0=-(S_1+S_2),\nonumber \\
&&2W_3=-S_1+S_2,\nonumber \\
&&2W_1=-(S_3+S_4),\nonumber \\
&&2U_2=-S_3+S_4.
\end{eqnarray}
Together with (3.24) this gives another representation of Killing vectors
(3.2--13). Identification of differential operators acting
on $z_i$ with the corresponding Killing vectors acting
on the initial target space variables leads to the system of partial
differential equations which is solved in the Appendix.
As a result we obtain the following explicit expression
for the matrix $Z$:
 \begin{equation}
\ Z=H^{-1}
\left(\begin{array}{ccc}
(1-z)(i\cal E \rm +1)-{\rm \Phi}^2 & \hspace{5mm} & i\cal E\it+z \\
i\cal E\it+z & \hspace{5mm} & (1+z)(i\cal E \rm -1)+{\rm \Phi}^2  \\
\end{array}\right),
\end{equation}
\begin{equation}
H=i{\cal E}z+(\Phi+1)^2.
\end{equation}
Here $z$ is the dilaton--axion (4.14), and
\begin{equation}
\Phi =u-zv,\quad
\cal E=\it f+i\chi-iv\rm \Phi.
\end{equation}
are two complex variables which may be regarded as generalization
of the Ernst potentials \cite{er} to dilaton--axion gravity.

Sigma--model of Sec. 2 can now be rewritten as a coset model for the
matrix $M$ (4.8). Let us introduce a matrix current
\begin{equation}
{\cal J}=\nabla  MM^{-1}.
\end{equation}
Substituting (4.8) into (4.29) we get
{\small
\[ \cal J=
\left(\begin{array}{rrr}
-P^{-1}\nabla P-P^{-1}\nabla QP^{-1}Q & \hspace{2mm} &
P^{-1}\nabla QP^{-1}\\
\nabla Q-QP^{-1}\nabla P-\nabla PP^{-1}Q-QP^{-1}\nabla QP^{-1}Q &\hspace{2mm} &
\nabla PP^{-1}+QP^{-1}\nabla QP^{-1}
\end{array}\right).\]
}
Using this matrix, three--dimensional sigma--model action (2.9)
can now be written as
\begin{equation}
 S=\int \left(-{\cal R}+\frac{1}{4}Tr({\cal J}^2)\right)\sqrt{h}d^3x,
\end{equation}
while the corresponding field equations read
\begin{equation}
\nabla {\cal J}=0,
\end{equation}
\begin{equation}
{\cal R}_{mn}=\frac{1}{4}Tr({\cal J}_m{\cal J}_n).
\end{equation}

So far we have constructed a standard matrix $\sigma$--model for the
coset $Sp(4, R)/U(2)$.
Remarkably, there exists more elegant representation of the same model
directly by $2\times 2$ matrices.
Indeed, from three $2\times 2$ matrix equations following from  (4.30)
\begin{equation}
\nabla \left(P^{-1}\nabla Q P^{-1}\right)=0,
\end{equation}
\begin{equation}
\nabla \left(QP^{-1}\nabla Q P^{-1}+\nabla P P^{-1}\right)=0 ,
\end{equation}
\begin{equation}
\nabla \left(\nabla Q -\nabla P P^{-1} Q- Q P^{-1}\nabla P
-QP^{-1}\nabla Q P^{-1} Q\right)=0 ,
\end{equation}
only two are independent. Using (4.33) and (4.34) one can present (4.35) as
\begin{equation}
P\nabla \left(P^{-1}\nabla Q P^{-1}\right)P=0,
\end{equation}
which directly follows from (4.33).

Introducing two $2\times 2$ matrix currents
\begin{equation}
{\cal J}_1=\nabla \it PP^{-1},\quad
{\cal J}_2=\nabla \it QP^{-1}.
\end{equation}
we can rewrite the equations of motion as
\[
\nabla {\cal J}_1+{{\cal J}_2}^2=0,
 \]
\begin{equation}
\nabla {\cal J}_2-{\cal J}_1{\cal J}_2=0,
\end{equation}
 \[
{\cal R}_{mn}=\frac{1}{2}Tr\left({\cal J}_{1m}{\cal J}_{1n}+
{\cal J}_{2m}{\cal J}_{2n}\right).
 \]
It can also be checked that
\begin{equation}
Tr \left( {\cal J}_1^2+{\cal J}_2^2\right)=\frac{1}{2}Tr{\cal J}^2 .
\end{equation}

Further simplification comes from the introduction of a complex
matrix dilaton--axion $Z$ instead of its real and imaginary parts
(from an explicit for of $Z$ (4.26) it is clear that real and imaginary parts
are more involved because of the complicated structure of the denominator).
Then instead of two real $2\times 2$ currents one can build
one complex matrix current
\begin{equation}
{\cal J}_Z=\nabla Z{(Z-\overline{Z})}^{-1},
\end{equation}
and cast the action into the following form
\begin{equation}
S=\int\frac{1}{2}\left(-{\cal R}+
2Tr{\cal J}_Z\overline{{\cal J}_Z}\right)\sqrt{h}d^3x .
\end{equation}
The correspoding equations read
\[
\nabla {\cal J}_Z={\cal J}_Z({\cal J}_Z-{\bar{\cal J}_Z})=0,
\]
\begin{equation}
{\cal R}_{mn}=2Tr{\cal J}_{Zm}{\bar{\cal J}_{Zn}}.
\end{equation}

The metric itself of the target space (2.9) can be presented
in terms of the matrix--valued
one--forms corresponding to the above currents
\begin{equation}
\omega=dMM^{-1},\quad \omega_1=dPP^{-1},\quad \omega_2=dQP^{-1},\quad
\omega_Z=dZ (Z-{\bar Z})^{-1},\quad
\end{equation}
by three alternative ways
\begin{equation}
{ds}^2=\frac{1}{4}Tr(\omega^2)= \frac{1}{2}Tr(\omega_1^2+\omega_2^2)=
2 Tr(\omega_Z {\bar \omega_Z}).
\end{equation}

Let us show that in terms of the complex variables
$z^\alpha=(z_1, z_2, z_3),\;z^{\bar\alpha}=({\bar z_1}, {\bar z_2}, {\bar
z_3})$,
where $z^{\bar\alpha}$ is a shorthand for ${\bar z^{\bar \alpha}}$, we
deal with a K\"ahler $\sigma$--model. In fact, writing an Hermitian metric
on the target space as
\begin{equation}
dl^2=K_{\alpha{\bar\beta}}dz^{\alpha} dz^{\bar\beta},
\end{equation}
we can construct a non--degenerate two--form
\begin{equation}
\Omega=\frac{i}{2}K_{\alpha{\bar\beta}}dz^{\alpha} \wedge dz^{\bar\beta}.
\end{equation}
One can easily check (using matrix dilaton--axion) that this form is closed:
\begin{equation}
d\Omega=0,
\end{equation}
what means that we have a K\"ahler manifold.


\section{Monopole-axion multicenter solution}
\renewcommand{\theequation}{5.\arabic{equation}}
\setcounter{equation}{0}

A representation of the  four--dimensional stationary
Einstein--Maxwell--dilaton--axion  system by a single complex
symmetric $2\times 2$ matrix opens new prospects for
integration of the system.  In particular, the inverse scattering transform
technique due to Belinskii and Zakharov \cite{bz} can be applied to
construct multisoliton solutions. Leaving this to  a subsequent publication,
we want to describe here some very simple (but apparently previously
unknown) solution  which is directly suggested by the above representation.

Consider equations (4.37), (4.38) written in terms of two currents related
to imaginary and real parts of $Z$.  A natural assumption allowing to
simplify the system drastically is that either real or imaginary part of $Z$
is a constant matrix. It is just the second assumption which
leads straightforwardly
to a new physically interesting solution. Thus, let us put
\begin{equation}
{\rm Im }Z=P=P_0={\rm const}.
\end{equation}
 From (4.37) it follows that  ${\cal  J}_1=0$ and the system (4.38) reduces to
\begin{equation}
\nabla{\cal J}_2=0,
\end{equation}
\begin{equation}
{\cal J}_2^2=0,
\end{equation}
and
\begin{equation}
{\cal R}_{mn}=\frac{1}{2} {\rm Tr} ({\cal J}_{2m}{\cal J}_{2n}).
\end{equation}

 From (5.2) we get a Laplace equation for the matrix $Q$
\begin{equation}
\nabla^2 Q=0,\quad \nabla^2=h^{-1/2}\partial_n(h^{mn}h^{1/2}\partial_m).
\end{equation}
This equation expresses solutions in terms of harmonic maps. Although
it is not difficult to treat more general case, we restrict ourselves here
by a one--parametric subspace $Q=Q[\lambda]$, where $\lambda$ is
some real scalar function on the 3--space. Then we get
\begin{equation}
\nabla^2 Q=Q^{''}(\nabla \lambda)^2 +Q^{'} \nabla^2 \lambda,
\end{equation}
where primes denote derivatives with respect to $\lambda$.
To satisfy this equation we simply take
\begin{equation}
Q^{''}=0,\quad \nabla^2 \lambda=0,
\end{equation}
i.e. choose $Q$ to be a linear function of $\lambda$:
\begin{equation}
Q=Q_0 +N\lambda,
\end{equation}
where $Q_0$ and $N$ are constant real symmetric matrices.

The second equation (5.3) is a constraint equation which imposes
the following condition on the matrix $N$
\begin{equation}
NP_0^{-1}N=0.
\end{equation}
It follows then that $N$ is degenerate, $\det N=0$. Finally the last
equation (5.4) tells us that the 3--space is Ricci--flat. Indeed,
\begin{eqnarray}
{\cal J}_{2m}{\cal J}_{2n}&=&Q_{,m} P_0^{-1}Q_{,n}P_0^{-1}\nonumber\\
&=& \lambda_{,m}\lambda_{,n}NP_0^{-1}NP_0^{-1},
\end{eqnarray}
so that the constraint (5.9) means that the Ricci tensor iz zero. Again we
take the simplest solution for the three--metric
$h_{mn}=\delta_{mn}$ (flat 3--space).

In view of the degeneracy of $N$ one can write
\begin{equation}
N=n n^T,\quad n=\left(\begin{array}{c}
n_1\\
n_2\\
\end{array}\right),
\end{equation}
where $n_1, n_2$ are real numbers. From (5.9) one gets
\begin{equation}
n^TP_0^{-1}n=0,
\end{equation}
which means that the ``metric'' $P_0^{-1}$ must have zero signature.

Now we specify the solution of the Laplace equation (5.7) on a flat 3--space
as follows
\begin{equation}
\lambda=4U,\quad U=\sum_i \frac{\mu_i}{|{\bf r}- {\bf r}_i|},
\end{equation}
where $\mu_i$ are real constants.
Denoting an asymptotic constant value of the matrix  dilaton--axion
as $Z_{\infty}$,
one can write the resulting solution as
\begin{equation}
Z=Z_{\infty}+\lambda N,
\end{equation}
what means that this solution can be generated from vacuum
by some $R$--transformation,
\begin{equation}
R=\lambda nn^T.
\end{equation}
 Choosing  $Z_{\infty}$ so that
\begin{equation}
f_{\infty}=-u_{\infty}=1,\quad \chi_{\infty}= \phi_{\infty}=v_{\infty}=0,
\end{equation}
one gets
\begin{equation}
P=-2\sigma_1,
\end{equation}
(Pauli matrix), and hence the desired signature property.
Now from (5.12) we obtain $n_1n_2=0$, and without loss of
generality one can choose $n_1=1, n_2=0$. It is a simple matter
now to write down a solution in physical terms
\begin{equation}
f=1, \quad u+1=-\chi=-\kappa=U.
\end{equation}
  This solution describes a system
of point--like monopoles carrying also axion charges of the same
strength and of opposite sign. Since the 3--space is flat,
the charges should be  in equilibrium
 (what is the case indeed: a magnetic repulsion is exactly
balanced by an axion attraction).  However, the 4--space is only locally flat.
To determine the four--metric we need to solve the equation
\begin{equation}
\nabla\times{\bf \omega}=-\nabla U.
\end{equation}
The solution corresponds to a lattice of Taub--NUT singularities
located at ${\bf r}= {\bf r}_i$:
\begin{equation}
ds^2=\left(dt-\omega_i dx^i\right)^2-d{\bf r}^2,
\end{equation}
where $d{\bf r}^2=dx^2+dy^2+dz^2$.
This is a massless multi--Taub--NUT solution whose emergence
for the present extremal monopole--axion system could be anticipated.

\section{Conclusion}
This work is a direct continuation of our previous papers \cite{gk}, \cite{g},
where the $U$--duality symmetry (1.1) was found in the $D=4,\;d=p=1$ theory.
In this case the symmetry group is $SO(2,3)$
which is isomorphic to the real symplectic group $Sp(4,R)$. In three
dimensions the system is described by six real scalar fields forming a coset
$Sp(4,R)/U(2)$. We have found a convenient parametrization of this coset
in terms of complex potentials generalizing Ernst potentials in the
Einstein--Maxwell theory. They can be combined into one complex symmetric
matrix (matrix dilaton--axion) which transforms under
$U$--duality via the matrix--valued $SL(2,R)$
group.  On the target space parametrized by three complex coordinates
(elements of this matrix) an Hermitian metric can be introduced which
generates a K\"ahler structure.

Complex representation is suggestive for new solution
generating techniques in the dilaton--axion gravity. A simple class of
solutions corresponding to constant imaginary part of the matrix
dilaton--axion is indicated, and a particular  solution is
found explicitly, which describes the system of magnetic/axion charges
in equilibrium. The relevant geometry can be viewed as massless
multi--Taub--NUT space--time. Our present formulation of the dilaton--axion
gravity also simplifies considerably an application to this theory of the
inverse scattering transform method. We intend to discuss this in more detailes
in a separate publication.

In higher--dimensional theories (including the full heterotic string
theory) $U$--duality group is no more isomorphic to any symplectic group.
However it is likely that at least some features of the formulation presented
here
for  $D=4,\; d=p=1$ theory will persist for higher values of $d,\;p$ as well.

\renewcommand{\theequation}{6.\arabic{equation}}
\setcounter{equation}{0}

\vskip3mm
 {\large \bf Acknowledgments}\\
One of the authors (D.V.G) is grateful to the International Center for
Theoretical Physics
Trieste, Italy, for hospitality during  the S--duality and Mirror Symmetry
Conference
while the final version of the paper was written. He thanks A. Sen  for
discussion
and C. Hull for clarifying some points. This work was supported in part by the
Russian
Foundation for Fundamental Research Grant 93--02--16977, and by the
International
Science Foundation and Russian Governement  Grant M79300.

\section{Appendix }
\renewcommand{\theequation}{A.\arabic{equation}}
\setcounter{equation}{0}
 Here we give detailes of the derivation of an explicit form of the
complex matrix $Z$ in terms of real target space variables.  First
of all we need an alternative representation of Killing vectors
through the holomorphic generators introduced in sec. 4. This can be achieved
by matching together  the correspondence rules
(3.24) and  (4.25)  :
\begin{eqnarray}
&&4K_g=-{(z_1-z_3+1)}^2\partial _1-{(z_2-z_3-1)}^2\partial _2+
(z_1-z_3+1)(z_2-z_3-1)\partial _3,\nonumber\\
&&4K_E=-{(z_1+z_3-1)}^2\partial _1-{(z_2+z_3+1)}^2\partial _2-
(z_1+z_3-1)(z_2+z_3+1)\partial _3, \nonumber\\
&&4K_{d_1}=-{(z_1+z_3+1)}^2\partial _1-{(z_2+z_3-1)}^2\partial _2-
(z_1+z_3+1)(z_2+z_3-1)\partial _3,\nonumber\\
&&4K_{d_2}={(z_1-z_3-1)}^2\partial _1+{(z_2-z_3+1)}^2\partial _2-
(z_1-z_3-1)(z_2-z_3+1)\partial _3,\nonumber\\
&&2K_s=({z_1}^2-{(z_3-1)}^2)\partial _1-({z_2}^2-{(z_3+1)}^2)\partial _2 +
(z_1+z_2+z_3(z_1-z_2))\partial _3,\nonumber\\
&&2K_{d_3}=-({z_1}^2-{(z_3+1)}^2)\partial _1-({z_2}^2-{(z_3-1)}^2)\partial _2 +
(z_1+z_2-z_3(z_1-z_2))\partial _3,\nonumber\\
&&2K_e=(1-{(z_1-z_3)}^2)\partial _1+(1-{(z_2-z_3)}^2)\partial _2+
(1+(z_1-z_3)(z_2-z_3))\partial _3,\nonumber\\
&&2K_{H_1}=-(1-{(z_1+z_3)}^2)\partial _1-(1-{(z_2+z_3)}^2)\partial _2+
(1+(z_1+z_3)(z_2+z_3))\partial _3 ,\nonumber\\
&&2K_m=({z_3}^2-{(z_1+1)}^2)\partial _1-({z_3}^2-{(z_2-1)}^2)\partial _2 +
z_3(z_2-z_1-2)\partial _3,\nonumber\\
&&2K_{H_2}=({z_3}^2-{(z_1-1)}^2)\partial _1-({z_3}^2-{(z_2+1)}^2)\partial _2 +
z_3(z_2-z_1+2)\partial _3.
\end{eqnarray}

Now  one can identify these operators with the initial Killing vectors
(3.2--13)
acting on the real target space variables. This gives ten triples  of partial
differential equations for three unknown complex functions $z_i$ in terms
of the derivatives
over the target space coordinates $f, \chi, u, v, \phi, \kappa$,
which can be solved via the step by step integration. The simplest
set comes from identification of the gravitational gauge Killing vectors
$K_g$:
\begin{eqnarray}
4\partial_{\chi }z_1&=&-{(z_1-z_3+1)}^2,\nonumber \\
4\partial_{\chi }z_2&=&-{(z_2-z_3-1)}^2,\nonumber \\
4\partial_{\chi }z_3&=&(z_1-z_3+1)(z_2-z_3-1).
\end{eqnarray}
This triple allows one to find an explicit dependence of the holomorphic
coordinates on $\chi$:
\begin{eqnarray}
z_1&=&{(b+1)}^2\zeta +c-1,\nonumber \\
z_2&=&{(b-1)}^2\zeta +c+1,\nonumber \\
z_3&=&(b^2-1)\zeta +c    ,
\end{eqnarray}
where  $\zeta ={(a+\chi)}^{-1}$. Here new variables $a, b, c$  depend
on five coordinates $f, u, v, \phi, \kappa$.

 At the second step we extract the $u$--dependence by identifying the
magnetic gauge generators:
\begin{eqnarray}
2{K}_m(z_1)&=&{z_3}^2-{(z_1+1)}^2 ,\nonumber \\
2{K}_m(z_2)&=&-{z_3}^2-{(z_2-1)}^2,\nonumber \\
2{K}_m(z_3)&=&z_3(z_2-z_1-2),
\end{eqnarray}
where action of the operator $K_m$ in terms of on real target space
coordinates is understood, Eq.(3.3). This results in relations
\begin{equation}
c=\gamma,\quad
b=\beta -\gamma u,\quad
a=\alpha +(2\beta -v)u-\gamma u^2,
\end{equation}
where $\alpha, \beta, \gamma$ are $\chi, u$--independent.

Explicit dependence on $v$ can be obtained from identification of the
electric gauge generators
\begin{eqnarray}
2{K}_e(z_1)&=&1-{(z_1-z_3)}^2,\nonumber \\
2{K}_e(z_2)&=&1-{(z_2-z_3)}^2,\nonumber \\
2{K}_e(z_3)&=&1+(z_1-z_3)(z_2-z_3).
\end{eqnarray}
Solving this system with account for previously found
interrelations one obtains
\begin{equation}
\gamma =\lambda,\quad
\beta =v+\nu,\quad
\alpha =2v+u(v+2\nu )-\lambda u^2+\mu.
\end{equation}
New variables $\lambda, \nu, \mu$ now depend only  on $f, \phi, \kappa$.

Further reduction comes from the system corresponding to $K_{d_1}$
\begin{eqnarray}
4{K}_{d_1}(z_1)&=&-{(z_1+z_3+1)}^2,\nonumber \\
4{K}_{d_1}(z_2)&=&-{(z_2+z_3-1)}^2,\nonumber \\
4{K}_{d_1}(z_3)&=&-(z_1+z_3+1)(z_2+z_3-1),
\end{eqnarray}
which gives
\begin{equation}
\lambda ={(l+\kappa )}^{-1},\quad
\nu =n\lambda ,\quad
\mu =m-n^2\lambda,
\end{equation}
with three functions $l, m, n$ of two variables $f, \phi$.
Now the explicit dependence on $\phi$ can be found from the $K_{d_3}$ triple
\begin{eqnarray}
2{K}_{d_3}(z_1)&=&{(z_3+1)}^2-{z_1}^2,\nonumber \\
2{K}_{d_3}(z_2)&=&{z_2}^2-{(z_3-1)}^2,\nonumber \\
2{K}_{d_3}(z_3)&=&z_1+z_2+z_3(z_2-z_1),
\end{eqnarray}
and reads
\begin{equation}
l=re^{-2\phi},\quad
n=pe^{-\phi}-1,\quad
m=q,
\end{equation}
where $r, p, q$ are functions of a single variable $f$.
Finally the scale transformation
\begin{eqnarray}
2{K}_s(z_1)&=&{z_1}^2-{(z_3-1)}^2,\nonumber \\
2{K}_s(z_2)&=&{(z_3+1)}^2-{z_2}^2,\nonumber \\
2{K}_s(z_3)&=&z_1+z_2+z_3(z_1-z_2),
\end{eqnarray}
completes determination of unknown functions up to
three constants of integration $r_c, p_c, q_c$
\begin{equation}
r=r_c,\quad
p=p_c f^\frac{1}{2},\quad
q=q_c f .
\end{equation}

Note, that in the above reduction it was sufficient to use only six ``simple''
Killing vectors in order to extract all functional dependence of $z_i$
on  $f, \chi, u, v, \phi, \kappa$. Two other Killing vectors
$K_{d_2}$  and $K_{H_1}$ now fix numerical constants
\begin{equation}
r_c^2=-1,\quad
p_c=0,\quad q_c=-r_c,
\end{equation}
so that only the sign of $r_c=\pm i$ remains unspecified. Since the
action (4.41) is invariant under complex conjugation, this sign
can be chosen arbitrarily, and we set $r_c=i$.

The remaining equations for the most involved Killing vectors $K_{H_2}, K_E$
will be automatically satisfied in view of the commutation relations
\begin{equation}
[K_{d_2}, K_{H_1}]=K_{H_2},\quad
[{K}_{H_1},{K}_{H_2}]=K_E.
\end{equation}

Therefore we have obtained the following explicit form of the functions
$a,b,c$:
\begin{equation}
c=z^{-1},\quad
b=v-(u+1)z^{-1},\quad
a=-if+(u+2)v-{(u+1)}^2z^{-1},
\end{equation}
where $z=\kappa +ie^{-2\phi}$. Substituting them into (A.3) and
introducing Ernst--type potentials (4.28) we arrive at
\begin{eqnarray}
&&z_1=\left(i{\cal E}-z-2(\Phi +1)\right)H^{-1}-1,\nonumber\\
&&z_2=\left(i{\cal E}-z+2(\Phi +1)\right)H^{-1}+1,\nonumber\\
&&z_3=\left(i{\cal E}+z\right)H^{-1},
\end{eqnarray}
where $H=i{\cal E}z+(\Phi +1)^2$.

The inverse relations are also useful
\begin{eqnarray}
&&z=-4W^{-1}(z_1+z_2-2z_3),\nonumber\\
&&{\cal E}=-4iW^{-1}(z_1+z_2+2z_3),\nonumber\\
&&{\Phi}=4W^{-1}(z_1-z_2+2)-1,
\end{eqnarray}
where
\begin{equation}
W= 4\left(z_3^2-(z_1+1)(z_2-1)\right).
\end{equation}
Substituting here $z_i$ from  (A.17) we find a simple identity
\begin{equation}
HW=16.
\end{equation}
\vskip1cm


\begin{thebibliography}{99}
\bibitem{gpr}  A. Giveon, M. Porrati, and E. Rabinovici, Phys. Rept.
{\bf 244} (1994) 77.
\bibitem{al} E. Alvarez, L. Alvarez--Gaume, and Y. Lozano, Nucl. Phys. B (Proc.
Suppl.)
{\bf 41}(1995)1.
\bibitem{nr} K. Narain, {\it Phys. Lett.} {\bf B169} (1986) 41;
K. Narain, H. Sarmadi, and E. Witten, {\it Nucl. Phys.} {\bf B279} (1987) 369.
\bibitem{odp}
G. Veneziano, Phys. Lett. {\bf B265} (1991) 287;  \\
K. Meissner and G. Veneziano, Phys. Lett. {\bf B267} (1991) 33; \\
M. Gasperini, J. Maharana and G. Veneziano, Phys. Lett. {\bf B272}
(1991) 277.
A. Sen, Phys. Lett. {\bf B271} (1991) 295; Phys. Lett. {\bf B274}
(1991) 34.
S.F. Hassan and A. Sen, Nucl. Phys. {\bf B375} (1992) 103.
A. Sen, Nucl. Phys. {\bf B404} (1993) 109.
\bibitem{sd}  A. Font, L. Iba\~nez, D. L\"ust, and F. Quevedo, {\it Phys.
Lett.} {\bf B249} (1990) 35;
S.J. Rey, {\it Phys. Rev.} {\bf D43} (1991) 526;
A. Sen, Nucl. Phys. {\bf B404} (1993) 109; Phys. Lett. {\bf 303B}
(1993); Int. J. Mod. Phys. {\bf A8} (1993) 5079; Mod. Phys. Lett. {\bf
A8} (1993) 2023;
J.H. Schwarz and A. Sen, Nucl. Phys. {\bf B411} (1994) 35; Phys.
Lett. {\bf 312B} (1993) 105.
\bibitem{s4} A. Sen, {\it Int. J. Mod. Phys.} {\bf A9} (1994) 3707.
\bibitem{gk}
D.V. Gal'tsov and O.V. Kechkin,  Phys. Rev. {\bf D50} (1994) 7394;
(hep--th/9407155).
\bibitem{eh}
J. Ehlers, in {\em Les Theories Relativistes de la
Gravitation}, CNRS, Paris, 1959, p. 275;
B.K. Harrison, Journ. Math. Phys. {\bf 9}, 1774 (1968).
\bibitem{g} D.V. Gal'tsov, Phys. Rev. Lett. {\bf 74} (1995) 2863,
(hep-th/9410217).
\bibitem{s3}  A. Sen, {\it Nucl. Phys.} {\bf B434}
(1995) 179, hep-th/9408083.
\bibitem{ms}  N. Marcus and J.H. Schwarz,
{\it Nucl. Phys.} {\bf B228} (1983) 145;
M. Duff and J. Lu, {\it Nucl. Phys.} {\bf B347} (1990) 394;
M. Duff and J. Rahmfeld, {\it Phys. Lett.} {\bf B345} (1995) 441,
hep-th/9406105.
\bibitem{ht}  C. Hull and P. Townsend, {\it Unity of Superstring
Dualities}, hep-th/9410167.
\bibitem{sh1} J.H. Schwarz, {\it Classical Symmetries of Some
Two-Dimensional Models}, CALT-68-1978, hep-th/9503078.
\bibitem{wi}
E. Witten, {\it String Theory Dynamics in Various Dimensions},
hep-th/9503124.
\bibitem{maz}
P.O. Mazur, Acta Phys. Polon. {\bf B14}, 219 (1983);
A. Eris, M. G\"urses, and A. Karasu, Journ. Math. Phys.
{\bf 25}, 1489 (1984).
\bibitem{gg} D.V. Gal'tsov, A.A. Garcia, in: Abstracts of Cornelius Lanczos
Int. Conf.,
Dec. 1993, North Carolina, p.100;  D.V. Gal'tsov, A.A. Garcia, and O.V.
Kechkin,
{\em Symmetries of the Stationary Einstein--Maxwell Dilaton Theory},
hep-th/9504155.
\bibitem{bk} I. Bakas, {\it Nucl. Phys.} {\bf B428} (1994) 374;
{\it Phys. Lett.} {\bf B343} (1995) 103.
\bibitem{mah}J. Maharana, {\it Hidden Symmetries of Two Dimensional
String Effective Action}, hep-th/9502001;
{\it Symmetries of the Dimensionally Reduced
String Effective Action}, hep-th/9502002.
\bibitem{bz}
V.A. Belinskii and V.E. Zakharov. Sov. Phys. JETP, {\bf 48},
985 (1978); {\bf 50}, 1 (1979);
 P. Breitenlohner and D. Maison, p. 276 in {\it
Explicit and Hidden Symmetries of Dimensionally Reduced (Super-) Gravity
Theories}.  Proceedings of the International Seminar on Exact Solutions of
Einstein's Equations.  Springer-Verlag (1983);
{\it Ann. Inst. H. Poincar\'e} {\bf 46} (1987) 215.
\bibitem{ki}
W. Kinnersley,  Journ. Math. Phys. {\bf 14}, 651 (1973);
{\bf 18}, 1529 (1977);
W. Kinnersley and D. Chitre, Journ. Math. Phys.
{\bf 18}, 1538 (1977); {\bf 19}, 1926, 2037 (1978).
\bibitem{s2} A. Sen, {\it Duality Symmetry Group of Two Dimensional
Heterotic String Theory}, preprint TIFR-TH-95-10, hep-th/9503057.
\bibitem{sh2}
J.H. Schwarz, {\it Classical Symmetries of Some
Two-Dimensional Models}, preprint CALT-68-1978, hep-th/9503078;
{\it Classical Duality Symmetries in Two Dimensions},
preprint CALT 68--1994, hep-th/9505170.
\bibitem{alk}
A. Kumar, K. Ray, {\it  Ehlers Transformations and String Effective Action},
preprint IP/BBSR/95-18, hep-th/9503154;
A.K. Biswas, A. Kumar, K. Ray, {\it Symmetries of Heterotic String Theory},
preprint IP/BBSR/95-51, hep-th/9506037;
I.R. Pinkstone, {\it Structure of dualities in bosonic string theory},
DAMTP R94-62, hep-th/9505147.
\bibitem{gz}
M.K. Gaillard and B. Zumino, Nucl. Phys. {\bf B193} (1981) 221.
\bibitem{hts}
G.T. Horowitz and A.A. Tseytlin, Phys. Rev. {\bf 51} (1995) 2896.
\bibitem{wzw}
E. Witten,  Phys. Rev. {\bf D44} (1991) 314.;
 S. B. Giddings, J. Polchinski and A. Strominger,
Phys. Rev. {\bf D48} (1993) 5748;  D. A. Lowe
and A. Strominger Phys. Rev. Lett.{\bf 73} (1994) 1468,
(hep-th/9403186); C. V. Johnson,   Phys. Rev. {\bf D50} (1994)
4032, hep-th/9403192.
\bibitem{gi}G.W. Gibbons, Nucl. Phys. {\bf B207} (1982) 337; G.W. Gibbons and
K. Maeda, Nucl. Phys. {\bf B298} (1988) 741; D. Garfinkle, G.T. Horowitz and
A. Strominger, Phys. Rev. {\bf D43}, (1991) 3140; R. Kallosh, A. Linde, T.
Ortin, A. Peet, A.
Van Proeyen, Phys.Rev. {\bf D46} (1992) 5278.
\bibitem{jm}
C.V. Johnson, {\it A Conformal Theory of a Rotating Dyon},
preprint PUPT--1524, McGill/95--01, hep-th/9503027.
\bibitem{sbh}
A. Sen, Nucl. Phys. {\bf 440} (1995) 421.
\bibitem{nk}
G. Neugebauer and D. Kramer, Ann. der Physik (Leipzig)
{\bf 24}, 62 (1969).
\bibitem{ggk}
D.V. Gal'tsov, A.A. Garcia, and O.V. Kechkin,
{\it Symmetries of the stationary Einstein--Maxwell--Dilaton--Axion
System} to be published in Journ. Math. Phys.
\bibitem{iw}
W. Israel and G.A. Wilson, Journ. Math. Phys. {\bf 13} (1972) 865.
\bibitem{bgm}
P. Breitenlohner, D. Maison, and G. Gibbons,
Comm. Math. Phys. {\bf 120}, 253 (1988).
\bibitem{ma}
D. Maison, Gen. Rel. and Grav. {\bf 10}, 717 (1979).
\bibitem{er}
F.J. Ernst, Phys. Rev. {\bf 167} (1968) 1175; {\bf 168} (1968) 1415;
I. Hauser and F. J. Ernst, Phys. Rev. {\bf D 20}, 362, 1783
(1979); Journ. Math. Phys. {\bf 21}, 1126, 1418 (1980).

\end{thebibliography}
\end{document}